\documentclass[onecolumn]{svjour3}

\usepackage[utf8]{inputenc}         % source files are utf8 encoded

\usepackage[T1]{fontenc}            % you always want this!

\usepackage{mathptmx}
\usepackage{graphics,graphicx}
\usepackage{amsmath}
\usepackage{amssymb}
\usepackage[usenames,dvipsnames]{xcolor}
\usepackage{xspace}
\usepackage{url}
\usepackage{algorithm}
\usepackage{algorithmic}
\usepackage{multirow}
\usepackage{listings}
\usepackage{color}
\usepackage[loose]{units}
\usepackage{booktabs}
\usepackage{todonotes}

\usepackage{bm}
\colorlet{shadecolor}{gray!25}

\usepackage{acronym}
\acrodef{DSL}{Domain-Specific Language}
\acrodef{FPGA}{Field Programmable Gate Array}

\newlength{\algrhswidth}
\newcommand{\algrhs}[1]{\hfill \parbox[t]{\algrhswidth}{#1}}

\usepackage{relsize}
\newcommand{\CPP}{C\nolinebreak[4]\hspace{-.05em}\raisebox{.3ex}{\relsize{-1}{++}}~}
\newcommand{\hipacc}{%\textsf{
  HIPA%\nolinebreak[4]
  \hspace{-0.1em}\textsuperscript{cc}}%}

\newcommand{\examultirow}[2]{\multirow{#1}{*}{\rotatebox[origin=c]{90}{\centering #2}}}

\newcommand{\Walberla}{\textsc{waLBerla}}
\newcommand{\exastencils}{\textsc{ExaStencils}}
\renewcommand{\div}{\operatorname{div}}

\newcommand{\refeqn}[1]{(\ref{#1})}

\colorlet{fhcolor}{ProcessBlue}

\lstdefinelanguage{scala}{
  morekeywords={abstract,case,catch,class,def,%
    do,else,extends,false,final,finally,%
    for,if,implicit,import,match,mixin,%
    new,null,object,override,package,%
    private,protected,requires,return,sealed,%
    super,this,throw,trait,true,try,%
    type,val,var,while,with,yield},
  otherkeywords={=>,<-,<\%,<:,>:,\#,@},
  sensitive=true,
  morecomment=[l]{//},
  morecomment=[n]{/*}{*/},
  morestring=[b]",
  morestring=[b]',
  morestring=[b]"""
}

\DeclareMathAlphabet{\mathpzc}{OT1}{pzc}{m}{it}

\journalname{}

\begin{document}

%\pretitle{Pretitle}
\title{A Scala Prototype to Generate Multigrid Solver Implementations for Different Problems and Target Multi-Core Platforms}
%\subtitle{Subtitle}
\titlerunning{A Scala Prototype to Generate Multigrid Solver Implementations}

\author{
	Harald K\"ostler \and
	Christian Schmitt \and \\
	Sebastian Kuckuk \and
	Frank Hannig \and \\
	J\"urgen Teich \and
	Ulrich R\"ude
}	

\institute{   Department Informatik, Technische Fakult\"at\\
							Friedrich-Alexander-Universit\"at Erlangen-N\"urnberg (FAU) \\
              \email{harald.koestler@fau.de}           %  \\
}


\maketitle

\begin{abstract}
Many problems in computational science and engineering involve partial differential equations and thus require the numerical solution of large, sparse (non)linear systems of equations. Multigrid is known to be one of the most efficient methods for this purpose. However, the concrete multigrid algorithm and its implementation highly depend on the underlying problem and hardware. Therefore, changes in the code or many different variants are necessary to cover all relevant cases.     
In this article we provide a prototype implementation in Scala for a framework that allows abstract descriptions of PDEs, their discretization, and their numerical solution via multigrid algorithms. From these, one is able to generate data structures and implementations of multigrid components required to solve elliptic PDEs on structured grids. 
Two different test problems showcase our proposed automatic generation of multigrid solvers for both CPU and GPU target platforms. 
\end{abstract}

\section{Introduction}
\label{sec:introduction}

Writing software for applications in Computational Science and Engineering (CSE) requires typically knowledge from different disciplines like physics, math, engineering, and computer science.
Today, a lot of widely used frameworks exist for different classes of CSE applications. However, when developing new applications or with a new generation of hardware often the frameworks have to be extended or even re-implemented.  
One solution to this problem is to provide a generic description of CSE algorithms, like already present in many frameworks, which can be transformed into optimized code via code generation techniques.

Examples for generation of mathematical code by using automatic tuning (optimization) techniques include ATLAS \cite{Whaley00automatedempirical} and the FFTW generator \cite{Frigo:1999:FFT:301618.301661} for small transformations.
Another example aimed at a broader application range, and thus a larger language definition, is FEniCS \cite{fenics:book}.
Being originally a library for finite element methods, it was extended to feature a Domain-Specific Language \ac{DSL} called UFL which is embedded in Python.
Julia \cite{2012arXiv1209.5145B} is centered around the multiple dispatch feature of object-oriented programming to provide distributed parallel execution and a mathematical functionality library.
Technically, it builds on a just-in-time compiler and also provides interface to calling \textit{Python} and \textit{C} functions.
Copperhead \cite{Catanzaro:2011:CCE:1941553.1941562} is a data-parallel language embedded in Python which targets CUDA and multicore CPUs using either Thread Building Blocks or OpenMP.
Pochoir \cite{tang2011pochoir}, Liszt \cite{devito2011liszt} and PATUS \cite{christen2011patus} are \acp{DSL} geared towards the description of stencil codes.
Whereas Pochoir generates \CPP{} code from descriptions embedded into \CPP{}, PATUS is written in Java and uses Coco/R as a parser generator and the Cetus framework, and Liszt is embedded into Scala and able to generate MPI, pthreads or CUDA code.

In the area of image processing, specialized approaches exist.
Examples include \hipacc{} \cite{MHTKE12a,MHTK12} and Halide \cite{Ragan-Kelley:2013:HLC:2491956.2462176}, both are \acp{DSL}s embedded into \CPP{}, allow the generation of code for different targets, among them CUDA and OpenCL. SPIRAL~\cite{puschel2004spiral,puschel2005spiral} generates platform-tuned implementations of digital signal processing (DSP) algorithms.

Our approach is in general similar to others with respect to the code generation techniques, however none of them so far has a scope as broad as the work presented. In other domains like

Currently, the CSE expert mainly focuses on parallel algorithms, implementation or extension of a framework, and tuning to a specific hardware platform, e.\,g.\ GPU. 
The users of the framework have to port the application to it, i.\,e., the model and the solution method have to be formulated in the language of the framework. If the framework already supports the
application, this means that just suitable modules have to be used, if not, the framework has to be extended. 
Thus, the user has to implement the application directly,
perhaps with the help of the framework developers. 

In contrast to a classical framework, which contains a collection of algorithms and applications and can be extended manually, we propose an approach to generate implementations that are adapted to a specific application and platform automatically. 
The code generation process itself has to work not only for a single application, but a certain class of applications (also called \textit{domain}) typically defined by the underlying algorithms. Note that in most cases frameworks will not become obsolete, but will be complemented by this approach, since they can be integrated in the generated code as external modules.

The key for success is of course the right choice of the class of applications. It must not be too large or too diverse.
In this article we focus on the domain of classical CSE applications that require the solution of elliptic PDEs via multigrid methods on regular grids.

In practice multigrid algorithms are implemented in various frameworks, e.\,g.\ in \Walberla{} for finite difference discretizations on
fully structured grids also supporting GPU clusters~\cite{KoestlerRitterHarbin10}, Boomer AMG \cite{Falgout99} for unstructured grids and general matrices, Peano~\cite{bungartz2010pde} that is based on space-filling curves, or DUNE~\cite{Bastian2008} that is a general software framework for solving PDEs.
Further examples of geometric multigrid solvers are found in~\cite{Adams00} for an unstructured Finite Element (FE) elasticity and plasticity problem and for a variable-coefficient Poisson problem with a proposed matrix-free distributed octree geometric multigrid algorithm in~\cite{sampath2010parallel} or
an algebraic multigrid solver on GPUs~\cite{haase2010parallel}. The widely used hypre\footnote{\url{http://www.llnl.gov/CASC/hypre/}} software package is a collection of high performance pre-conditioners and solvers including geometric and algebraic multigrid algorithms that scales to more than $100,000$ cores~\cite{baker2012scaling}.  

The reason for such a variety of different multigrid software packages is that the multigrid method has to be tuned not only to the specific (linear) system but also the parallel hardware in order to run efficiently.  
However, adopting the multigrid solver for different problems and target hardware setups leads to a high implementation effort and can be a time consuming process. Additionally, a detailed domain knowledge is required. Within the project \textit{Advanced Stencil-Code Engineering} (\exastencils)\footnote{\url{http://www.exastencils.org/}} we try to collect this knowledge and use it to automatically generate highly parallel multigrid implementations for specific problems~\cite{koestlerkuckukparco13,koestlerapelhistencils13,kronawitterhistencils13}. It started in 2013 and is supported by the German Research Foundation (DFG) through the Priority Programme 1648 \textit{Software for Exascale Computing (SPPEXA)}. 

After restricting the class of applications, we thus still have several degrees of freedom like the type of hardware, the dimension of the problem, the exact PDE, or the concrete components for the multigrid solver. These degrees of freedom are collected in a so-called \textit{feature model} \cite{batory2005featuremodels}.

Our vision is that users can specify a problem at a high level of abstraction. 
The domain expert brings in algorithmic knowledge and is usually a CSE expert.
The mathematician sets constraints to the algorithmic components or provides knowledge to estimate the (problem-dependent) efficiency for certain algorithms.
Software specialists implement the code generation framework,
and the hardware specialist may provide problem-independent tuning of the implementation to a certain target platform.

The feature model then allows to build up a global optimization problem to find the optimal implementation for a certain problem, where algorithms, their components and parameters, and possible tuning strategies to a specific hardware are selected automatically from a list of possibilities. The search space for the optimization problem is constrained by domain knowledge, either specified by the domain experts, or found based on machine learning techniques \cite{siegmund2012predicting}.
Note that an automation of the above mentioned path to generate code for a whole CSE application is an ambitious goal that one can reach only step by step.

A review of different technologies for the implementation of \acp{DSL}~\cite{SKKHT14} led us to the choice of Scala for the implementation of the code generation framework. This paper presents a first code generation prototype written in Scala and is structured as follows: In Section~\ref{sec:abstract} we introduce a multi-layered approach for the abstract description of CSE applications, Section~\ref{sec:scala} provides implementation details for our prototype, and Section~\ref{sec:exp} shows first numerical results obtained with generated C++ or CUDA code for two model problems that are chosen such they require quite different components and algorithms.

\section{Abstract Description of CSE Applications\label{sec:abstract}}

\begin{figure}[ht!]
\begin{center}
\includegraphics[width=.7\linewidth]{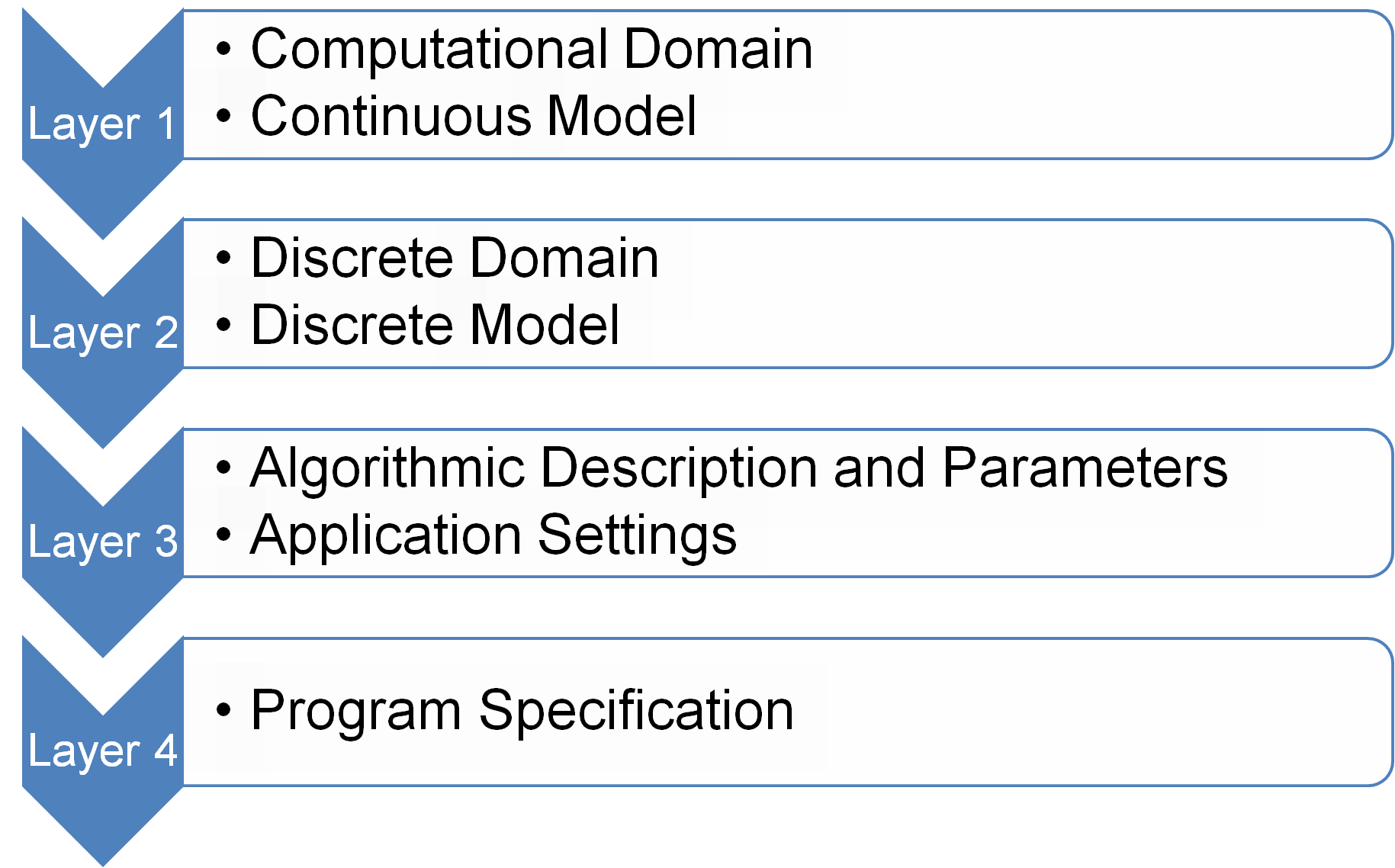}
\caption[Partitioning of a cell-centered grid into four sub-grids]{Different layers for describing a prototypical CSE application.}
\label{fig:applevels}
\end{center}
\end{figure}

The process to describe a prototypical CSE application is summarized in Figure~\ref{fig:applevels}. In the following, all layers are explained along two examples.  

In general, one can either describe the whole application on all layers in detail and then generate the full code, or one can build upon existing software and only replace performance critical parts like the multigrid solver by generated code. The latter is usually much easier to do. Note, however, that there is no guarantee that this approach scales very well, since other parts of the existing code can become the bottleneck for performance and portability then. 

As a showcase, we have implemented a first prototype in Scala\footnote{\url{http://www.scala-lang.org}}. It enables us to generate from abstract descriptions on all layers of describing a prototypical CSE application, depicted in Figure~\ref{fig:applevels}, either CUDA or C++ code for geometric multigrid solvers on regular grids in 2D and 3D.
We mainly choose Scala, because of the built-in support for parser combinators~\cite{odersky2008programming} and the possibility to benefit from object-oriented and also functional programming concepts, what helps to write code transformations in a compact form. %~\citeown{koestlerschmitthistencils13}.
The abstract descriptions are formulated in external domain-specific languages (DSLs)~\cite{ghosh2010dsls}, i.\,e., they are not embedded in another language. Context-sensitive grammars for them can be specified in a notation similar to Extended Backus–Naur Form (EBNF)~\cite{garshol2003bnf}.  

%{\textit TODO: Although the final DSLs' looks will be most likely similar to \LaTeX{} on Layers 1 and 2, Matlab\footnote{\url{http://www.mathworks.de}} on Layer 3, and Scala on Layer 4, we start with a simpler syntax and simple semantics on each Layer in our first prototype and only provide rudimentary DSLs stored in one file per Layer.}\fh{Ich weiß zwar in welche Richtung der Satz geht, aber ich würde auf keinem Fall mit \LaTeX{}, Matlab, etc. vergleichen, da irreführend.}

\subsection{Computational Domain and Continuous Model (Layer 1)}

\paragraph{Poisson's equation:} On layer 1 according to Figure~\ref{fig:applevels} the computational domain is fixed. As one simple test case, we consider Poisson's equation as an elliptic PDE that can be described by  
\begin{subequations} \label{eq:homodiffusion}
\begin{eqnarray}  
 \Delta u &=& f \quad \text{in} \; \Omega \\
   u  &=& 0 \quad \text{on} \; \partial \Omega \quad 
\end{eqnarray}
\end{subequations}
where we assume homogeneous Dirichlet boundary conditions.

\paragraph{DSL: Continuous problem and domain} The continuous problem from Eq.~\refeqn{eq:homodiffusion} on the domain $\Omega = [0,1]^2$ on Layer 1 can be described by the following DSL code:
\begin{verbatim} 
Domain d = [0,1] x [0,1]

Function f = 0
Unknown solution = initrandom
Operator Lapl = Laplacian

PDE pde { Lapl(solution) = f } 
PDEBC bc { solution = 0 }

Accuracy = 12
\end{verbatim}
On this Layer the user is able to define and name a domain, e.\,g., changing it to $[0,1] x [0,1] x [0,1]$ would result in solving a 3D instead of a 2D problem. \textit{Accuracy = 12} sets the desired physical distance between two grid points to $2^{-12}$ in order to determine the number of grid points in each dimension for discretization on Layer 2. % discretization error
The solution is initialized randomly. Furthermore, a function can be provided for the
right hand side $f$. 

Descriptions of all other Layers can be generated automatically from information on Layer 1 and by adding internal domain knowledge. However, it is possible to adapt them also after the default versions are created. 

\subsection{Discrete Domain and Discrete Model (Layer 2)}

%{\bf TODO: delete the next sentence?} First, the domain $\Omega$ is split into quadrangles that are then regularly refined using the HHG approach.  
Eq.~(\ref{eq:homodiffusion}) is discretized by finite differences. 
This results in a sparse linear system of equations $A^h u^h = f^h$
%\begin{equation}
%A^h u^h = f^h \; , \quad \sum_{j \in \Omega^h} a_{ij}^h u_j^h = f_i^h \, ,i \in \Omega^h \label{symsystem}
%\end{equation} 
with discretization matrix $A^h \in \mathbb{R}^{N \times N}$, vector of unknowns $u^h \in \mathbb{R}^N$ 
and right hand side (RHS) vector $f^h \in \mathbb{R}^N$ on a discrete grid $\Omega^h$. 
$N$ denotes the degrees of freedom or number of unknowns in the linear system.
In our case $A$ can be stored in a compact, matrix-free form using a constant 
5-point stencil. 

\paragraph{DSL: Discrete problem and domain} On Layer 2, the description of the default discretization of the above problem is:
\begin{verbatim} 
Fragments f1 = Regular_Square

Discrete_Domain d levels 12 { 
  xsize = 4096 
	xcoarsefac = 2
  ysize = 4096  
	ycoarsefac = 2
} 

field<Double,1>@nodes f  
field<Double,1>@nodes solution  
stencil<Double,FD,2>@nodes Lapl  
\end{verbatim}

A number of decisions had to be made here, either directly based on domain knowledge, or later based on domain-specific optimization. First, the computational domain is partitioned into fragments.   
Then the grid sizes can be derived from the accuracy on Layer 1 in this simple case. Further defaults are the underlying data type, here double precision floating point numbers, the number of components per grid point within a field, a node-based location of the grid points, and a second order finite difference discretization. %The numbers after the data type denote that it is a scalar PDE problem and not a system of PDEs. 
Note that vectors are mapped to fields and sparse matrices to stencils.

\subsection{Algorithmic Description and Parameters (Layer 3)}
\label{sec:MG}

In order to solve the above discrete (non)linear systems approximately, we apply a multigrid method~\cite{Brandt77,Hackbusch85}.
As an iterative solver for large, sparse linear systems its major advantage 
is that it has an asymptotically optimal complexity of $\mathcal{O}(N)$~\cite{Briggshenson00,Trottenbergoosterlee01,Griebel99}.

The multigrid idea is based on two principles:
the smoothing property, i.\,e., that classical iterative methods like Gauss-Seidel (GS) are able 
to smooth the error after very few steps, and the coarse grid principle, 
i.\,e., that a smooth function on a fine grid can be approximated satisfactorily on a grid with 
less discretization points.
Multigrid combines these two principles into a single iterative solver.
The smoother reduces the high frequency error components first, and then the low
frequency error components are approximated on coarser grids, prolongated
back to the finer grids and eliminated there.
This leads to recursive algorithms which traverse between fine and coarse grids in a grid
hierarchy. 
One multigrid iteration, the so-called \textit{V-cycle}, is summarized in 
Algorithm~\ref{Dis:AlgCS}. 

\begin{algorithm}
\caption[Multigrid correction scheme]{Recursive V-cycle: $u_h^{(k+1)} =
V_h(u^{(k)}_h,A^h,f^h,\nu_1,\nu_2)$}
\begin{algorithmic}[1]
\IF{coarsest level} \STATE solve $A^h u^h = f^h$ by a (parallel) direct solver or by CG iterations

\ELSE

\STATE $\bar{u}^{(k)}_h =
\mathcal{S}^{\nu_1}_h(u_h^{(k)}, A^h,f^h)$
\algrhs{\COMMENT{presmoothing}}

\STATE $r^h = f^h -
A^h \bar{u}^{(k)}_h$
\algrhs{\COMMENT{compute residual}}
\STATE $r^H = R r^h$
\algrhs{\COMMENT{restrict residual}}

\STATE $e^H =
V_H(0,A^H,r^H,\nu_1,\nu_2)$
\algrhs{\COMMENT{recursion}}

\STATE $\tilde{u}^{(k)}_h =
\bar{u}^{(k)}_h + P e^H$
\algrhs{\COMMENT{prolongate error and do coarse grid correction}}

\STATE $u^{(k+1)}_h =
\mathcal{S}^{\nu_2}_h(\tilde{u}_h^{(k)},
A^h,f^h)$ \algrhs{\COMMENT{postsmoothing}}
\ENDIF
\end{algorithmic} \label{Dis:AlgCS}

\end{algorithm}

Two successive grid levels are denoted by $\Omega^h$ and $\Omega^H$
and the following components are used in the 2D multigrid Poisson solver:
\begin{itemize}
\item a Jacobi (Jac) or a Red-Black Gauss-Seidel (RBGS) Smoother $\mathcal{S}^{\nu_1}_h,\mathcal{S}^{\nu_2}_h$ with
            $\nu_1$ pre- and $\nu_2$ postsmoothing steps, 
\item direct coarse grid approximation for the operator $A^h$, and 
\item a bilinear interpolation operator $P$ in order to prolongate the error to the next finer grid level and its transpose as restriction $R$.
\end{itemize}

%Note that performing parallel multigrid scaling experiments on the simple Poisson problem is one of the most challenging problems, since the ratio between computation and communication is very poor for an efficient parallel execution, since only few arithmetic operations are required.

\paragraph{DSL: Algorithmic components and parameters} On Layer 3 mainly default multigrid components and parameters are set (compare to Algorithm~\ref{Dis:AlgCS}). Note that for real-world CSE applications, input parameters and other algorithmic components are also prescribed on this Layer. 

\begin{verbatim} 
mgcomponents { 
  smoother = GaussSeidel 
  interpolation = interpolatecorr  
  restriction = Restrict 
  coarsesolver = GaussSeidel 
  cycle = VCycle 
} 

mgparameter { 
  nlevels = 7  // number of multigrid levels 
  restr_order = 2 // order of restriction operator
  int_order = 2 // order of interpolation operator
  ncoarse = 10 // number of coarse grid solver iterations
  nprae = 2 // number of pre-smoothing steps 
  npost = 1 // number of pre-smoothing steps 
  iters = 10 // maximum number of V-cycles
  omega = 1.0 // smoother parameter
} 
\end{verbatim}
%   accuracy = 8 // threshold for residual reduction

For each of the components the name of a corresponding function in the pseudo code implementation on Layer 4 is provided. With additional information, that the order of the restriction operator is 2, the restriction stencil from Layer 2 can now defined to be full weighting, the interpolation to be bilinear. The decisions on this Layer can be supported by a performance model including Local Fourier Analysis (LFA)~\cite{Wienands05} predictions for the multigrid convergence rates and domain knowledge. 

\subsection{Pseudo Code (Layer 4)}

Layer 4 offers an interface to add algorithms or links to modules from an external framework. After the code on this Layer has been generated from previous Layers, one is able to formulate own functions and classes in a Scala-like syntax close to concrete code. 
The application specialist can provide a short description of the application, which will become the main function in the generated code. The CSE expert can add new numerical algorithms, which work on the discrete domain and can access stencils and fields defined on Layer 2. One may also prescribe for each function, on which platform it should be executed. Basic language elements can be used, e.\,g., local variables, statements, expressions, several kinds of loops, and simple I/O like printing to the standard output. 

For the above test problem, all multigrid components are automatically generated using domain knowledge. The application is found in the main function, furthermore the V-cycle, the smoother, and the transfer operator functions are shown below. 
\begin{verbatim}
def cpu Application ( ) : Unit 
{  
  decl res0 : Double = L2Residual ( 0 ) 
  decl res : Double = res0 
  decl resold : Double = 0 
  print ( 'startingres' res0 ) 
  repeat up 10 
    resold = res 
    VCycle ( 0 ) 
    res = L2Residual ( 0 ) 
    print ( 'Residual:' res 'residual reduction:' (res0/res) ) 
  next  
}  

def cpu VCycle ( lev:Int  ) : Unit 
{ 
  if coarsestlevel { 
    repeat up ncoarse 
      GaussSeidel ( lev) 
    next  
  } else { 
    repeat up nprae 
      GaussSeidel( lev) 
    next  
	  Residual ( lev ) 
	  Restrict ( (lev+1) f[(lev+1)] Res[lev]) 
	  set( (lev+1) solution[(lev+1)] 0) 
	  VCycle (lev+1) 
	  interpolatecorr( lev solution[lev] solution[(lev+1)] ) 
    repeat up npost 
      GaussSeidel ( lev ) 
    next  
  } 
} 

def cpu GaussSeidel ( lev:Int ) : Unit  
{ 
    loop innerpoints level lev order rb block 1 1 
      solution = solution[lev] + (inverse( diag(Lapl[lev]) ) * omega 
			           * ( f[lev] - Lapl[lev] * solution[lev] ) ) 
    next  
}  

def cpu Restrict ( lev:Int coarse:Array fine:Array) : Unit  
{ 
    loop innerpoints level coarse order lex block 1 1  
      coarse = RestrictionStencil * fine | ToCoarse  
    next  
}  

def cpu interpolatecorr( lev:Int uf:Array uc:Array ) : Unit 
{ 
    loop innerpoints level uf order lex block 1 1  
      uf += transpose(RestrictionStencil) * uc | ToFine  
    next  
}  
\end{verbatim}

%\caption{V-cycle. compare to graphical and algorithm!}

Two important language features shown above are the loop concept and the matrix-vector product. The \textit{loop} keyword corresponds to a (nested) for-loop construct, which iterates over parts or the whole discrete computational domain (grid) in a certain manner. The first modifier specifies the part of the grid, currently \textit{allpoints}, \textit{innerpoints}, or \textit{boundarypoints}, \textit{level} on which grid level the loop iterates, \textit{order} the order of traversal through the grid points, and \textit{block}, if a point-wise or block-wise update is done. Inter-grid transfers are triggered by the additional modifiers \textit{ToCoarse} or \textit{ToFine} at the end of the statement within a loop. A matrix-vector (stencil-field) product is usually also found inside a loop over (parts of) the grid, where one requires data from neighboring grid points, when applying the stencil, depending on its shape and size. This is often the most expensive operation and thus has to be implemented efficiently. 

What is missing in this example are explicit calls to communication routines necessary for the generation of MPI-parallel code. Basically, they are treated just like usual function calls, where the function name is \textit{communicate} and the arguments are the field and the grid level, on which ghost layers should be exchanged. Low level optimizations like vectorization or blocking are not yet supported and will be done internally via code transformations and not specified explicitly in the DSL. Layer 4 has to be adapted, if our approach is combined with a new external framework in order to provide an interface to it. 

%\subsection{Target-specific information} 
%\begin{verbatim}
%Hardware cpu {
  %bandwidth = 60
  %peak = 118
  %cores = 4
%}
%
%Node {
  %sockets = 1  
%}
%
%Cluster {
  %nodes = 1
  %networkbandwidth = 10
%}
%\end{verbatim} 

%Information about peak performance (in GFLOP/s) and bandwidth (in GB/s) is important for the internal performance model, the type of hardware triggers, if C++ or CUDA code has to be generated, e.\,g., in the above specification, a platform is modeled with a single cluster ... If there is more than one core available on CPU, OpenMP support is enabled, and in case of more than one node, MPI support is enabled.

In addition to the problem, the user also lists basic facts about the target platform in a separate file.
The features of the prototype that can be set on different Layers are summarized in Table~\ref{tab:generator}. Currently, we are building up an automatic test suite such that it becomes easier to test all possible feature combinations. %MPI support will be provided by S. Kuckuk.

 \begin{table}
\centering
\caption{\label{tab:generator} A feature model for the first Scala prototype. Features in bold font have to be specified by the application expert, all others can be derived from these. } % measured cycles/FLOP ratio 
\begin{tabular}{lcl}
\toprule\noalign{\smallskip}
 Feature & Layer & Values   \\   
\noalign{\smallskip}\midrule\noalign{\smallskip}
\textbf{Computational domain} & 1 &  UnitSquare, UnitCube  \\
\textbf{Operator} & 1 &  Laplacian, ComplexDiffusion \\
\textbf{Boundary conditions} & 1 &  Dirichlet, Neumann  \\
Location of grid points & 2 &  node-based, cell-centered  \\
Discretization & 2 &  finite differences, finite volumes  \\
Data type & 2 &  single/double accuracy, complex numbers  \\
Multigrid smoother & 3 &  $\omega$-Jacobi, $\omega$-Gauss-Seidel, red-black variants  \\
Multigrid inter-grid transfer & 3 &  constant and linear interpolation and restriction  \\
Multigrid coarsening & 3 &  direct (re-discretization)  \\
Multigrid parameters & 3 &  various  \\
Implementation & 4 &  various code optimization strategies \\ % CG?
\textbf{Platform} & Hardware &  CPU, GPU  \\
Parallelization & Hardware &  serial, OpenMP \\
\noalign{\smallskip}\bottomrule
\end{tabular} 
\end{table}

\section{Code Generation Prototype \label{sec:scala}}

\paragraph{Structure}

\begin{figure}[ht!]
\begin{center}
\includegraphics[width=.98\linewidth, height=.87\textheight]{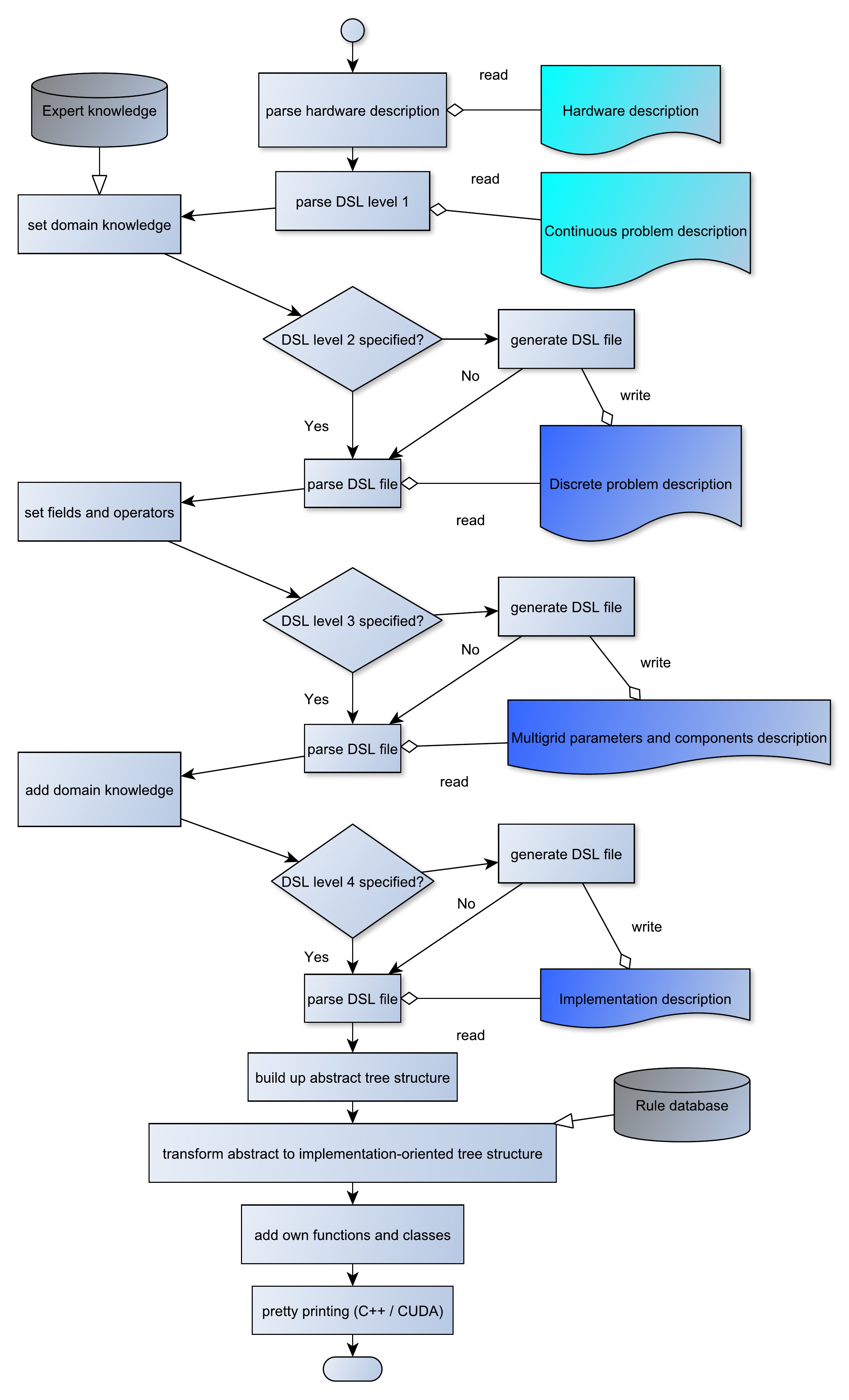}
\caption[Partitioning of a cell-centered grid into four sub-grids]{General structure of Scala prototype. DSL descriptions on different Layers are parsed, transformed into
an implementation-oriented representation, and then pretty-printed to C++ or CUDA source code.}
\label{fig:scalastructure}
\end{center}
\end{figure}

\begin{figure}[ht!]
\begin{center}
\includegraphics[width=.35\linewidth, height=.6\textheight]{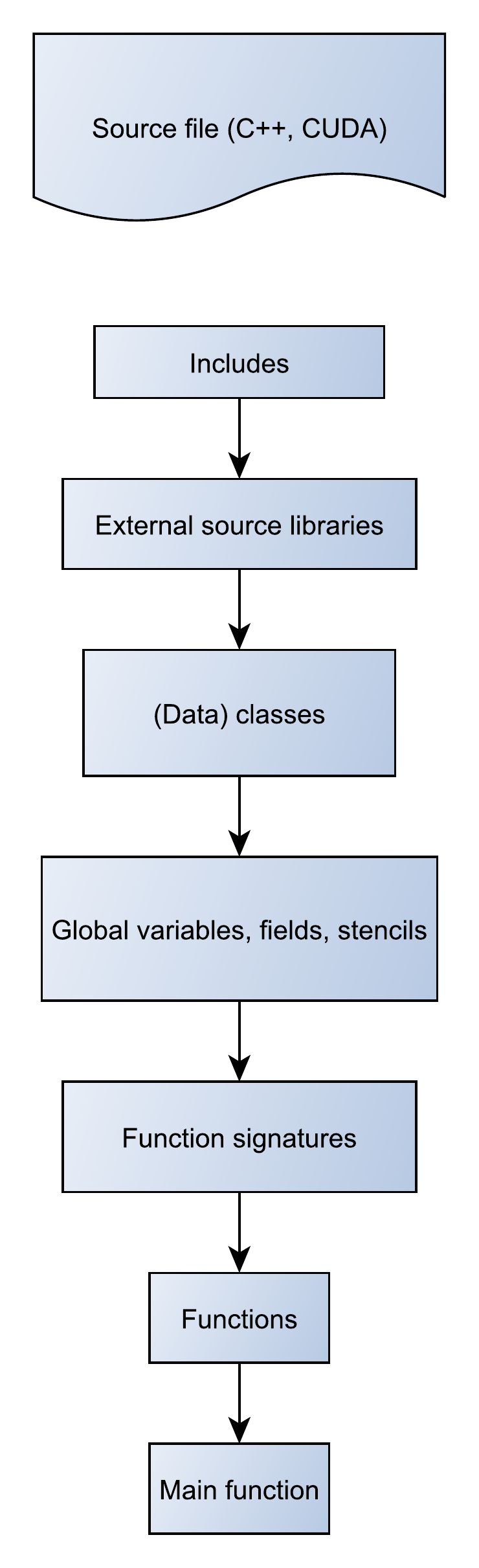}
\caption[Partitioning of a cell-centered grid into four sub-grids]{Structure of the pretty-printed source file.}
\label{pix:scalasource}
\end{center}
\end{figure}

\begin{figure}[ht!]
\begin{center}
\includegraphics[width=.94\linewidth]{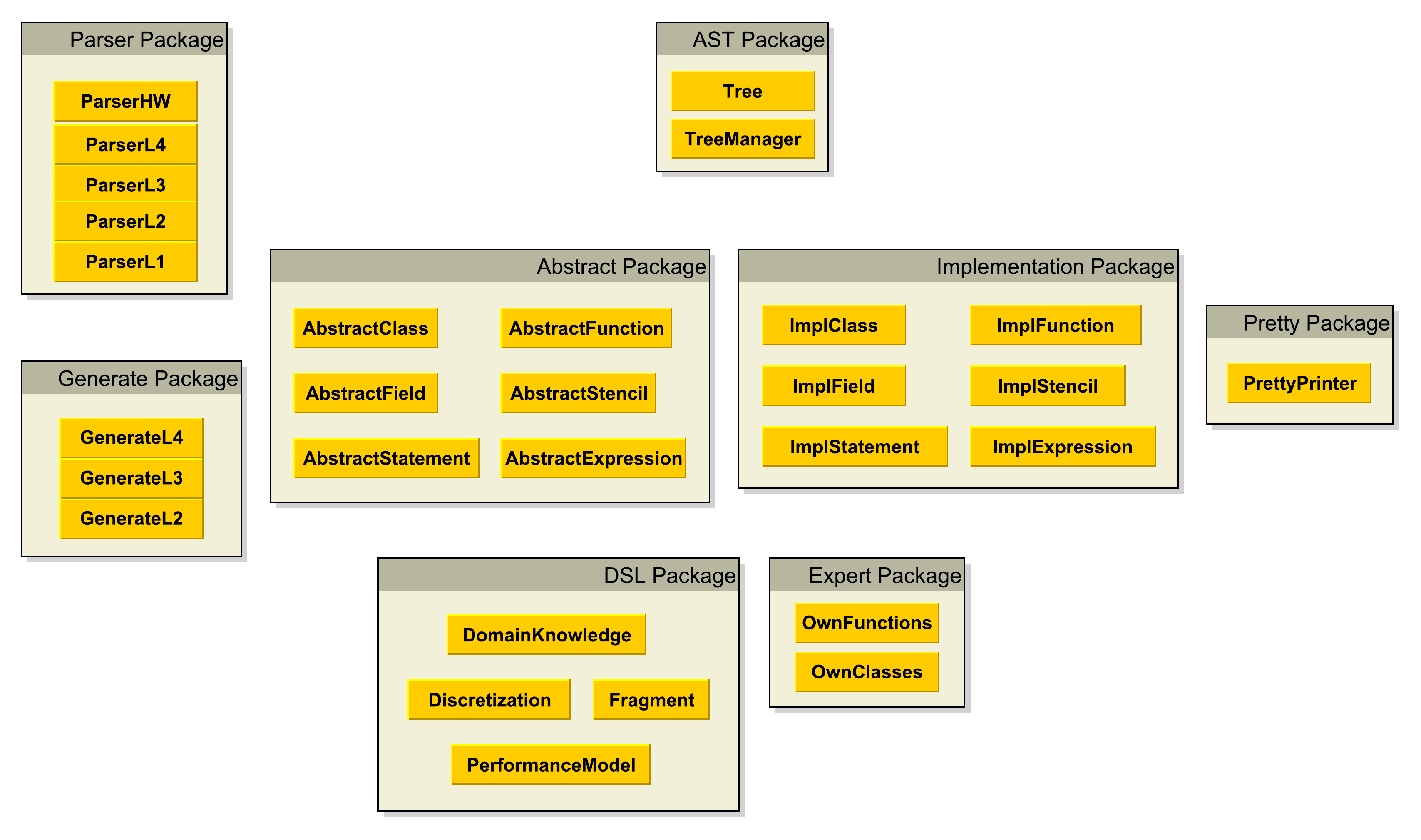}
\caption[Partitioning of a cell-centered grid into four sub-grids]{Packages of the Scala prototype.}
\label{fig:scalapackages}
\end{center}
\end{figure}

With the above information of the target platform and information from all higher Layers, one can finally generate CPU or GPU code as summarized in Figure~\ref{fig:scalastructure}. 
In Figure~\ref{pix:scalasource} one finds the structure of the pretty-printed source file.
Our Scala prototype implementation consists of several packages listed in Figure~\ref{fig:scalapackages}.

%The parsers on each Layer is realized by Scala parser generators. 
%Parser: different scala parser for each Layer
%        L1+HW: descriptions by user (problem)
Parsers for all Layers are found in the \textit{Parser Package}.
The \textit{Generate Package} uses domain knowledge formulated as rules to decide on Layer 2, how to discretize the problem, on Layer 3, which multigrid components and parameters are suitable, and on Layer 4, how the multigrid components are implemented. These rules will be collected in a rule data base later in order to enable an automatic learning process and an easy way to extend the domain knowledge.

%Generate: L2: how to discretize problem (domain, operator, unknown, f)
%          L3: which multigrid components + parameters are used? (and other parameters !)
%          L4: algorithms

%Tree: stores list of classes, functions, (ghost)fields, stencils, definitions, callgraph
%      has transformation functions

%\paragraph{Representation of domain knowledge}

The class \textit{DomainKnowledge} stores global data from all Layers, global variables, fields, and stencils. %(ghost)fields, stencils (one per level). 
It contains, e.\,g., global rules for determining array sizes and helper functions for index mapping. Furthermore, it defines how stencil operations and boundary conditions are implemented. % (stencils, boundary fields)

\paragraph{Representation of computational domain and discretization}

A \textit{fragment} represents the geometry of the computational domain. It contains primitive classes, like faces, edges, and vertices. An example for the regular square is found next. 
          
\begin{minipage}{\textwidth}
\begin{lstlisting}[language=Scala]
  var fragments: ListBuffer[Fragment] = ListBuffer()

  def initfragments() {
    var v: List[Vertex] = List()
    var e: List[Edge] = List()
    var f: List[Face] = List()

    if (DomainKnowledge.fragment_L2.get._2.equals("Regular_Square")) {
      v = List(new Vertex(ListBuffer(0.0, 0.0)), new Vertex(ListBuffer(0.0, 1.0)), 
			         new Vertex(ListBuffer(1.0, 0.0)), new Vertex(ListBuffer(1.0, 1.0)))
      e = List(new Edge(v(0), v(1)), new Edge(v(0), v(2)), 
			         new Edge(v(1), v(3)), new Edge(v(2), v(3)))

      f = List(new Face(List(e(0), e(1), e(2), e(3)), v))

      fragments += new Fragment(f, e, v)
    }
  }
\end{lstlisting}
\end{minipage}

In later parallel simulations, a fragment is represented as a local block, the computational domain is split into fragments of the same shape and size, and each process owns one or several fragments.
The handling of boundary conditions and the parallel data exchange via ghost layers is also based on fragments, where a mapping, e.\,g., for copying data to or from boundary primitives like edges into data buffers, is described by an affine coordinate transformation.

%For discretization simple finite differences are supported, the finite element local stiffness matrices can be assembled using the COLSAMM library\footnote{\url{http://www10.informatik.uni-erlangen.de/People/Alumni/jochen/.www/colsamm.html}}, and for mehrstellen discretization we plan to generate the stencil entries via an approach described in~\cite{HeisigBA13}. % (refer to Daniel's student).

\paragraph{Code Transformations}

Parsers on each Layer create for each entry in the abstract description corresponding objects of one of the classes found in the \textit{Abstract Package}. These objects are transformed into objects of classes from the \textit{Implementation Package} that can already pretty-print themselves to C++ or CUDA code.  
As an example the Scala classes for fields in both packages are

\begin{minipage}{\textwidth}
\begin{lstlisting}[language=Scala]
class AbstractField(val name: String, val datatype: String, val location: String)

class ImplField(val name: String, val datatype: String, 
                val sizex: Int, val sizey: Int, val sizez: Int, val addpoints: Int)
\end{lstlisting}
\end{minipage}

The Scala classes for stencils look similar. For functions and classes it is more complex, here one has to construct an abstract syntax tree (AST) for each of them. 

\begin{minipage}{\textwidth}
\begin{lstlisting}[language=Scala]
class AbstractFunction(fname: String, location: String, rettype: String, 
                       paramlist: List[Param], stmts: List[AbstractStatement]) {
    def transform: ListBuffer[ImplFunction] = { return ... }
}

class ImplFunction(fname: String, location: String, rettype: String, 
                   paramlist: ListBuffer[ParameterInfo], 
									 bodylist: ListBuffer[ImplStatement]) {
    def toString : String = { return ... }							
}
\end{lstlisting}
\end{minipage}

\begin{figure}[ht!]
\begin{center}
\includegraphics[width=.78\linewidth]{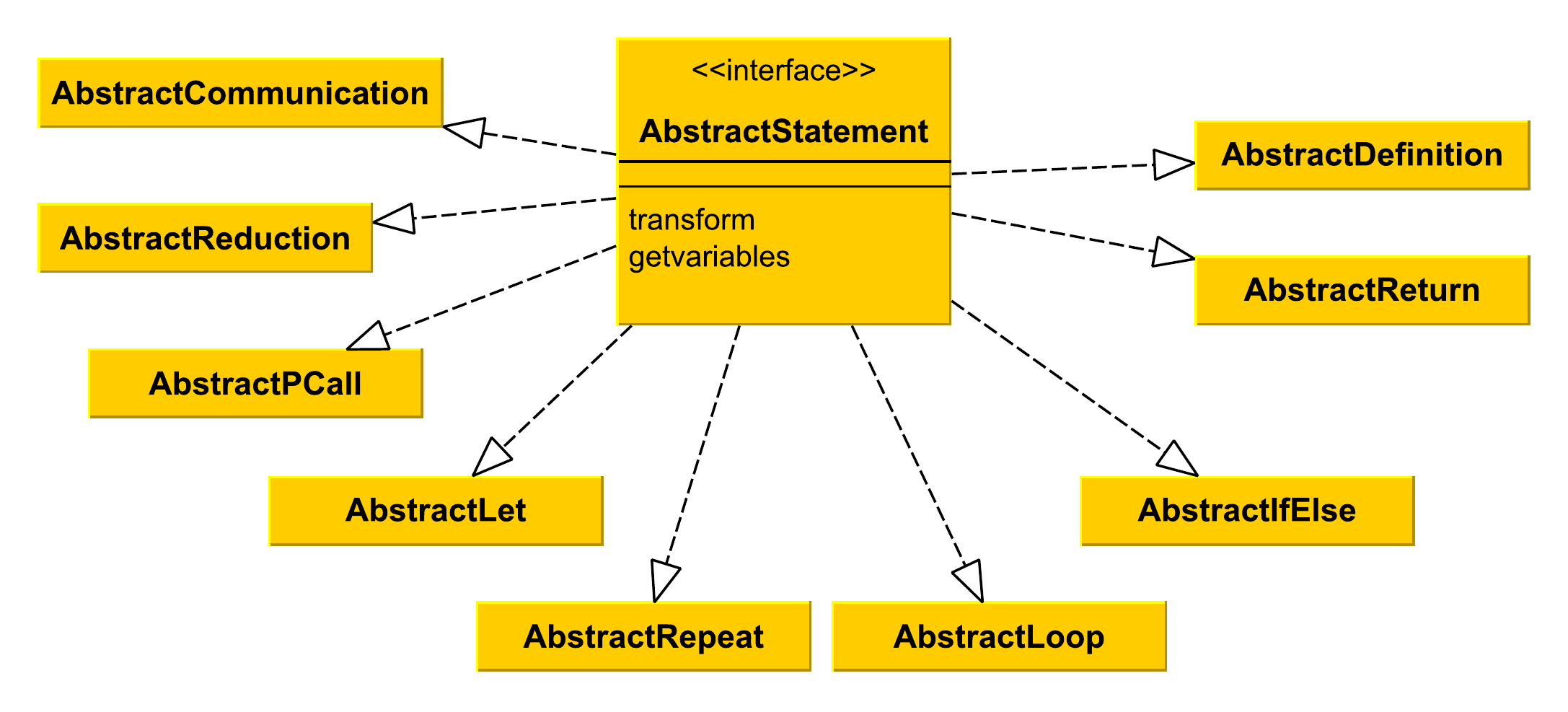}
\caption[Partitioning of a cell-centered grid into four sub-grids]{Class hierarchy for abstract statements.}
\label{fig:scalaabstractstate}
\end{center}
\end{figure}

\begin{figure}[ht!]
\begin{center}
\includegraphics[width=.65\linewidth]{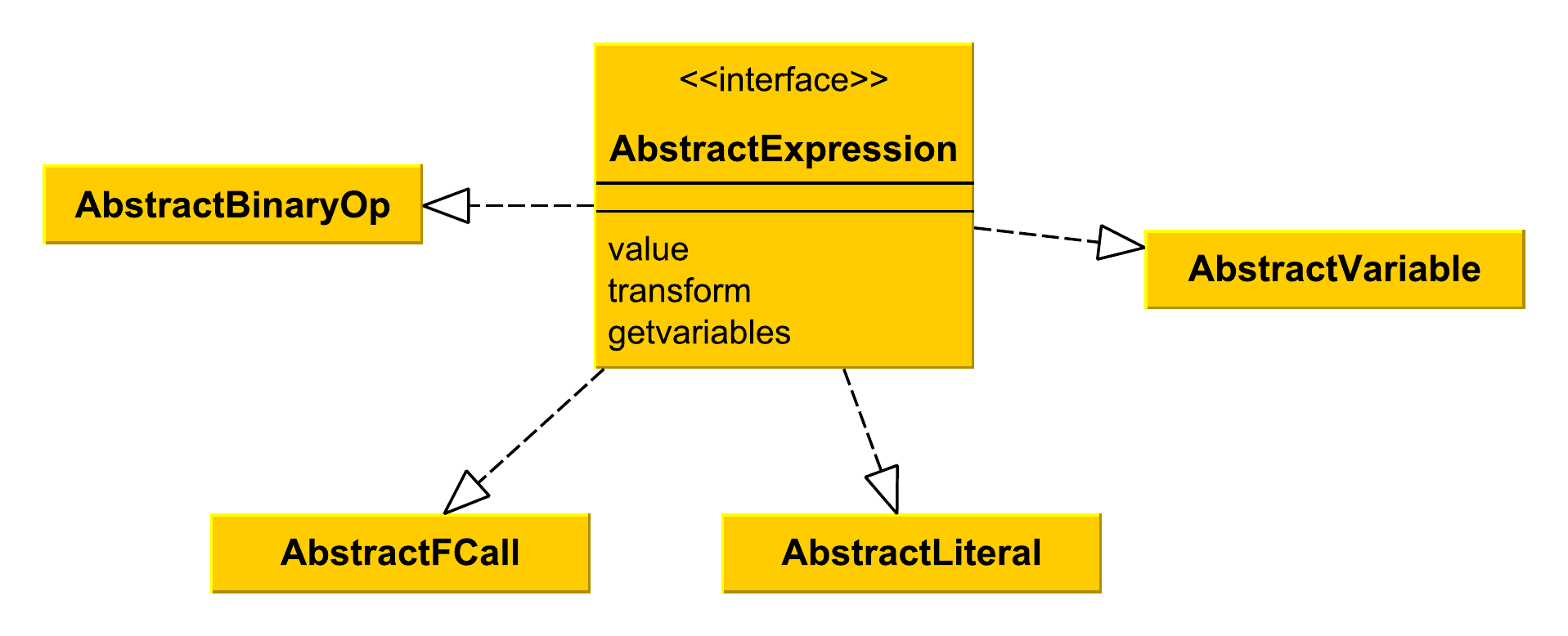}
\caption[Partitioning of a cell-centered grid into four sub-grids]{Class hierarchy for abstract expressions.}
\label{fig:scalaabstractexpr}
\end{center}
\end{figure}

\begin{figure}[ht!]
\begin{center}
\includegraphics[width=.82\linewidth]{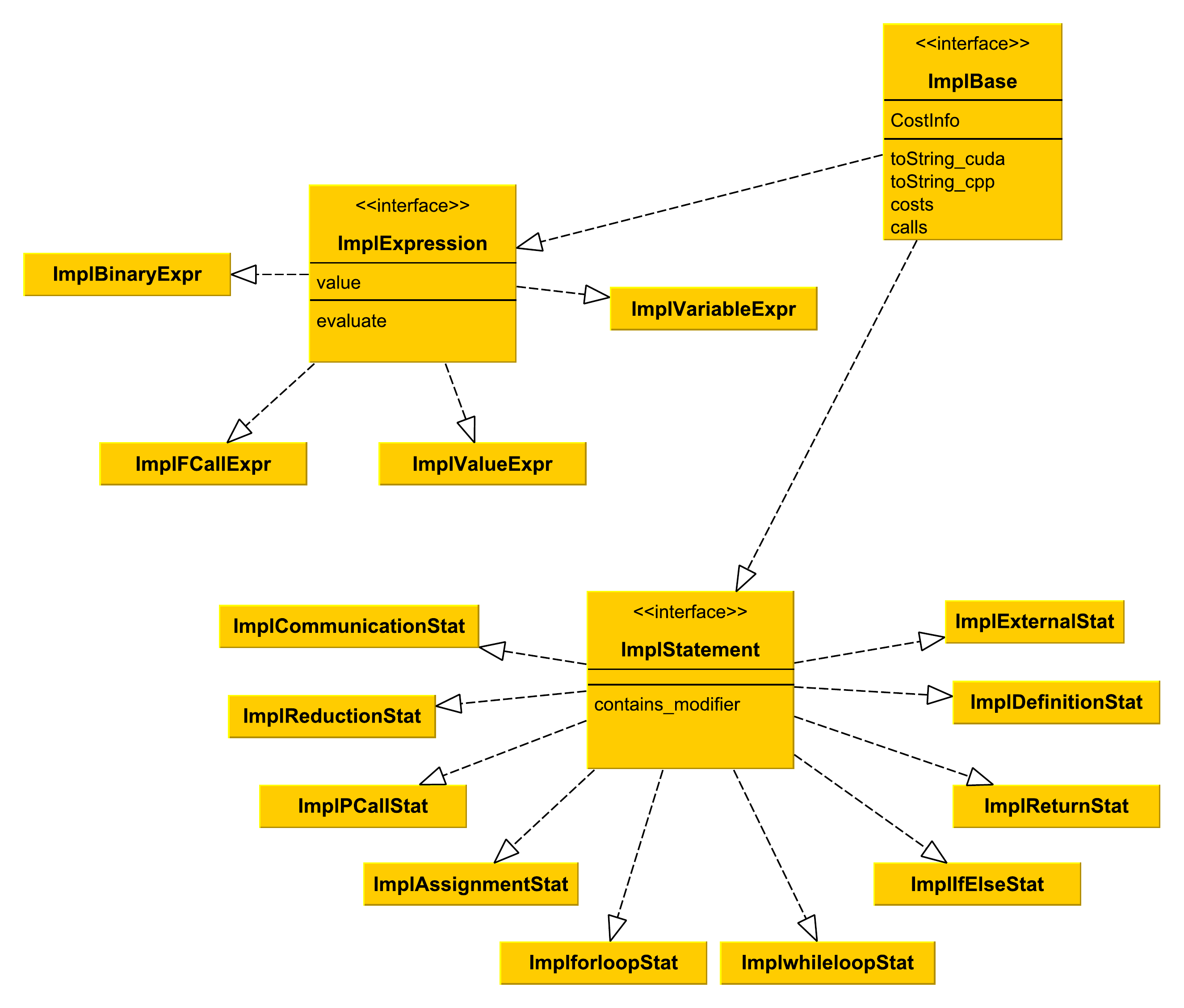}
\caption[Partitioning of a cell-centered grid into four sub-grids]{Class hierarchy for implementation-oriented statements and expressions.}
\label{fig:scalaimpl}
\end{center}
\end{figure}

Besides the function name, the location, i.\,e., CPU or GPU, the return type, the list of parameters, and the list of statements in the function body are specified. The routine \textit{transform} models the internal code transformations that involve, in general, information from all Layers and domain knowledge. For completeness, in Figure~\ref{fig:scalaabstractstate} all possible abstract statements, and in Figure~\ref{fig:scalaabstractexpr} all possible abstract expressions are depicted, which can occur in the AST. The implementation-oriented statements and expressions after code transformation are found in Figure~\ref{fig:scalaimpl}. % TODO: Note that high-level code transformations fct->fct or fine granular like stat->stat   

The \textit{AST Package} stores lists of objects for all defined classes, functions, fields, and stencils.
Currently, a more general code transformation framework is implemented, where one can apply also external transformations and it is possible to traverse the tree more than once. 

In the \textit{Expert Package} predefined classes for fields and stencils are contained including information about data layout and data access, the Stencil class also has member functions for matrix-vector operations, which are inlined later on. %: access + layout, convolution (Au,Ru,Iu)!
Furthermore, there are special functions, e.\,g., for initializing and cleaning up MPI, OpenMP, or CUDA, allocating fields and stencils, boundary treatment, and copy to or from MPI buffers.
%OwnFunctions: Main: init MPI/OpenMP/CUDA, allocate fields,stencils, init/set up problem, run application, clean up 
%              boundary treatment, copy to/from buffers

%level 2: AbstractStencil: mapping operator -> stencil knowledge
%transformations on different levels, fine granular or high level
%TODO: e.\,g.\ loop: red black GS

A routine found in the \textit{Pretty Package} writes the generated source string to a file that can be compiled directly.

\paragraph{GPU Support}

In principle, an abstract function, specified on Layer 4 to be located on GPU, is transformed into a CPU interface function and a GPU kernel function. The first sets up necessary information about GPU kernel block and grid sizes and then calls the GPU kernel. The computations are performed within the GPU kernel. Note that one has to assure that data is also present in GPU memory, i.\,e., GPU memory must be allocated before on GPU, and data transfers between CPU and GPU via the relatively slow PCIe bus can be necessary. Therefore, it typically makes no sense to switch too often between CPU and GPU kernels, but to transfer data once to GPU at the beginning, then do all computations on GPU, and transfer at the end the results back to CPU.

\section{Experimental Results} \label{sec:exp}

Currently, our Scala prototype is already capable of creating implementations for several different feature combinations, where the problem, the computational domain, and the platform may be varied. As a consequence, discretization, multigrid components and parameters, and thus the overall implementation changes significantly.

As test problems Poisson's equation (see Eq.\refeqn{eq:homodiffusion}) and complex diffusion are considered. Both can be formulated within our multi-layered DSL.

Complex diffusion is motivated by a simplified time-dependent Schr\"odinger equation. One can model a nonlinear isotropic complex diffusion process by the time-dependent PDE
\begin{eqnarray}  \label{eq:compldiffusion}
 \div \left( g \left( Im(u) \right) \nabla u \right)  &=& u_t
\end{eqnarray}
with Neumann boundary conditions, initial condition $u(0) = u^0$, and time $t = 0$.
$Im(u)$ denotes the imaginary part of $u$ and the complex diffusivity function is given by
\begin{equation}
g \left( Im(u) \right) = \frac{e^{i \theta}}{1 + \left( \frac{Im(u)}{k \theta} \right)^2} .
\end{equation} 
with small angle $\theta$ and scaling parameter $k > 0$. Complex diffusion is used, e.\,g., for denoising of images~\cite{gilboa2004,honigman2006}.

For discretization, averaged finite differences (corresponding to finite volumes) in space and an implicit Euler scheme in time are applied. 

The arising (non)linear system of equations is solved in each time step via a Full Approximation multigrid Scheme (FAS)~\cite{Trottenbergoosterlee01}, where the non-linearity is treated by lagged diffusivity. The transfer operators are cell-centered restriction and constant interpolation.
Note that one can change the multigrid V-cycle description on Layer 4 to the FAS scheme~\cite{Dietrich:2010:AWSEC10} easily.

%Hardware specification, multigrid components and parameters, refer to reference implementation from paper?

%\begin{table}
%\centering
%\caption{\label{tab:generatorresults} Measured runtimes in seconds for one V(2,2)-cycle for different feature combinations. Default settings are a serial run on one CPU core, red-black $\omega$-Gauss-Seidel with $\omega=1.0$, and double precision. In case of Jacobi smoother $\omega=0.8$. The number of unknowns is $N = 4095^2$ (node-based) resp. $N = 4096^2$ (cell-centered) in 2D, and $N = 255^3$ in 3D. Direct coarsening down to less than three unknowns per direction on the coarsest grid is applied. } % measured cycles/FLOP ratio 
%\begin{tabular}{llllc}
%\hline\noalign{\smallskip}
%Computational & Platform & PDE & Settings & runtime \\ % & estimate (IO) & estimate (FLOPs)   \\   
%Domain & & & & in s\\ 
%UnitSquare & CPU & Poisson &  & 0.7 \\ % & 0.28  & 0.12   \\
%UnitSquare & CPU & Poisson & Jacobi & 1.0 \\ %& 0.26 (0.08) & 0.02  \\
%UnitSquare & CPU & Poisson & OpenMP 4 threads, Jacobi & 0.3 \\ %& 0.28 (0.03) & 0.02  \\
%UnitCube  & CPU & Poisson &  & 1.1 \\ %& 0.45 & 0.19  \\
%UnitSquare & GPU & Poisson &  & 0.05 \\  % TODO: explain difference to paper!
%UnitSquare & CPU & ComplexDiffusion & Jacobi  & 32.1  \\
%UnitSquare & CPU & ComplexDiffusion & OpenMP 4 threads, Jacobi & 12.8   \\
%UnitSquare & GPU & ComplexDiffusion & single precision, Jacobi & 0.09  \\  % TODO: explain difference to paper!
%\end{tabular} 
%\end{table}

\begin{table}
\begin{center}

\begin{tabular}{ccc rrrrrr}
    \toprule
&&& \multicolumn{4}{c}{CPU} & \multicolumn{2}{c}{GPU}\\
\cmidrule{4-9}
&&& \multicolumn{2}{c}{1 Core} & \multicolumn{2}{c}{8 Cores} & \multicolumn{2}{c}{256 Threads}\\
\cmidrule{4-9}
&&& Double & Float & Double & Float & Double & Float\\
\midrule
\examultirow{5}{Laplacian} & \multirow{2}{*}{2D} &  Jacobi &         546   & 1.17 & 3.50 & 3.50 & 6.42  & 9.10\\
\cmidrule{3-9}
& &                                             Gauss-Seidel &       453   & 1.16 & 2.42 & 3.24 & 3.41  & 4.72\\
\cmidrule{2-9}
& \multirow{2}{*}{3D} &                             Jacobi &         608   & 1.08 & 2.60 & 3.00 & 6.83  & 10.31\\
\cmidrule{3-9}
& &                                         Gauss-Seidel &           608   & 1.08 & 2.60 & 3.01 & 4.22  & 5.96\\
\midrule
\examultirow{5}{Complex Diff.} & \multirow{2}{*}{2D} & Jacobi &      9,235 & 1.02 & 2.81 & 2.87 & 23.99 & 33.58\\
\cmidrule{3-9}
&&                                          Gauss-Seidel &           7,504 & 1.11 & 2.18 & 2.29 & 11.62 & 16.42\\
\cmidrule{2-9}
& \multirow{2}{*}{3D} &                         Jacobi &             8,799 & 1.15 & 2.77 & 2.91 & 15.94 & 39.28\\
\cmidrule{3-9}
&&                                          Gauss-Seidel &           9,048 & 1.80 & 2.60 & 2.87 & 10.28 & 24.19\\
\bottomrule
\end{tabular}

\caption[Partitioning of a cell-centered grid into four sub-grids]{\label{tab:generatorresults}Measured runtimes in milliseconds for one V(2,2)-cycle for different feature combinations. In case of Jacobi smoother $\omega=0.8$. The number of unknowns is $N = 4095^2$ (node-based) resp. $N = 4096^2$ (cell-centered) in 2D, and $N = 255^3$ resp. $N = 256^3$ in 3D. Direct coarsening down to less than three unknowns per direction on the coarsest grid is applied.}
\end{center}
\end{table}
 
Table~\ref{tab:generatorresults} summarizes some preliminary runtime results for generated codes using different feature combinations.
The experiments are conducted on a quad-core Intel Xeon Processor E5-1620 v2 running at 3.7 GHz and achieving a maximal peak memory bandwidth of 59.7 GB/s and 118.4 GFLOP/s peak performance, %3.7 GHz * one adder one multiplier ALU (2) * AVX vectorization (4) * cores (4) = 118.4 GFLOPs
and an NVIDIA GeForce GTX 680 achieving 192.2 GB/s and 3.1 TFLOP/s (single precision) respectively 128.8 GFLOP/s (double precision). 
As operating system Windows 7 is used, furthermore Visual Studio 2012 and the CUDA 5.5 compilers.

In case of Poisson, the $L_2$-norm of the residual is reduced by a factor of more than $10^8$ within 5 V(2,2)-cycles for all tests,
for complex diffusion by a factor of $10^5$. %(end $0.3$ convergence rate)
Note that the generated code tries to avoid manual problem-specific optimizations as far as possible, because the selection of suitable code transformations to obtain the most efficient implementation will be part of the global feature value optimization problem. %is currently not platform-optimized at all. This point will be addressed by S. Kronawitter in future. 
Complex diffusion on CPU is especially slow, where \textit{std::complex$<$double$>$} from the C++ standard library is used as built-in data type. For the GPU version a field with two components, one for the real and one for the imaginary part, is allocated and own device functions for complex arithmetic are provided.
A hand-tuned implementation for 2D Poisson takes about \unit[0.008]{s} on GPU in single precision~\cite{koestler13HDR}, with the same settings a hand-tuned complex diffusion takes about \unit[0.037]{s}. Optimizations in these implementations include, e.\,g., vectorization, change of data layout, simplification of index calculations and other computations, loop unrolling, and blocking techniques.

\section{Future Work}
\label{sec:conclusion}

We are currently generalizing the code transformation framework and re-defining the \acp{DSL} Layers in order to provide a clearer division of semantics that is available on each Layer.
Another goal is to ease collaboration between developers of the framework by providing more abstractions and interfaces.

To allow further optimizations towards the target hardware, an expressive hardware description language will need to be designed and implemented into the framework.
In the light of modern supercomputers becoming increasingly heterogeneous, we will need such a flexible approach to generate optimal code.
Examples for the settings that need to be described are CPU cache hierarchies, architectures of accelerators such as GPUs or \acp{FPGA} or the communication characteristics of cluster nodes.

Currently, the language definitions are lacking a concept for the handling of communication via MPI, resulting in the generated code being single-node only.
We can imagine many cases where it is not appropriate to let the compiler choose the data that is to be exchanged or the timing when this will happen.

The presented prototype currently does not allow for context-sensitive transformations, which is a huge drawback in the generation of specialized code variants, e.\,g., when applying optimizations for different target devices such as CPU or GPU.

The presented prototype already allows the selection of the data type used for calculations, i.\,e., double vs.\ float.
This, however, showed the need for a cleaner internal type system and a working type replacement strategy that were not considered in the first place.
Such an approach will be especially useful when dealing with specialized data types such as complex numbers or vector data types to enable further optimizations like SIMD extensions.
As can be seen from the performance evaluation, particularly the CPU code needs to be improved.
For easier optimization, we plan to switch to plain C for the computation kernels.

Once the code generation framework is completed, it will be possible to generate thousands of different multigrid implementations already out of feature combinations listed in Table~\ref{tab:generator}. The global optimization problem to find the best implementation for a certain application with respect to accuracy of the solution and time to compute the solution is in general infeasible to solve, of course. 
Our approach will be based on the feature model and first define, which parts of the optimization problem involve functional quantities, e.\,g., optimal multigrid parameters, and which parts involve non-functional quantities, e.\,g., the type of discretization or smoother \cite{koestlerapelhistencils13}. 
One idea is to optimize functional quantities based on performance models, LFA predictions, domain knowledge, and sample measurements. %The resulting accuracy of the solution together with runtime and convergence rate estimates or measurements can be used as fitness function within a multi-objective heuristic algorithm that optimizes (non-functional) feature strings.  

%Clearly, the above work is far off the traditional CSE applications with respect to the used algorithms. But it performs a simulation of the game, where in each time step the player chooses his next action that has to fulfill several constraints. The same now holds for the CSE expert when developing code. On each Layer several choices have to be made under certain restrictions and earlier decisions influence later ones.
%This domain knowledge about feature combinations and interactions can be extended by machine learning techniques in order to reduce the global search space for the multi-objective genetic algorithm.
%It will be also interesting to have the option to do multi-objective optimization, since optimizing only for minimal runtime may not lead to the best result, if the estimates are not very accurate. In this case it can be beneficial to optimize for a minimal runtime together with a good convergence rate resulting in non-domination Pareto fronts.

%machine learning for non-functional quantities like type of smoother, type of discretization, type of domain decomposition, memory layout
%for non-functional optimization we require, e.\,g., heuristic genetic algorithm.

\section{Acknowledgments}

\exastencils is funded by the German Research Foundation (DFG) as part of the Priority Programme 1648 (Software for Exascale Computing). The first funding period runs from January 2013 to December 2015. 
%We gratefully acknowledge the Gauss Centre for Supercomputing (GCS) for providing computing time through the John von Neumann Institute for Computing (NIC) on the GCS share of the supercomputer \Juqueen{} at J\"ulich Supercomputing Centre (JSC). GCS is the alliance of the three national supercomputing centres HLRS (Universit\"at Stuttgart), JSC (Forschungszentrum J\"ulich), and LRZ (Bayerische Akademie der Wissenschaften), funded by the German Federal Ministry of Education and Research (BMBF) and the German State Ministries for Research of Baden-W\"urttemberg (MWK), Bayern (StMWFK) and Nordrhein-Westfalen (MIWF).

%
%
%\bibliographystyle{spmpsci}
%\bibliography{Literature,cv_harald_deutsch,LiteraturHabil}

\begin{thebibliography}{10}
\providecommand{\url}[1]{{#1}}
\providecommand{\urlprefix}{URL }
\expandafter\ifx\csname urlstyle\endcsname\relax
  \providecommand{\doi}[1]{DOI~\discretionary{}{}{}#1}\else
  \providecommand{\doi}{DOI~\discretionary{}{}{}\begingroup
  \urlstyle{rm}\Url}\fi

\bibitem{Adams00}
Adams, M., Demmel, J.: Parallel multigrid solvers for 3d unstructured element
  problems in large deformation elasticity and plasticity.
\newblock In: International Journal for Numerical Methods in Engineering, pp.
  48--8 (2000)

\bibitem{baker2012scaling}
Baker, A., Falgout, R., Kolev, T., Yang, U.: Scaling hypre's multigrid solvers
  to 100,000 cores.
\newblock In: High-Performance Scientific Computing, pp. 261--279. Springer
  (2012)

\bibitem{Bastian2008}
Bastian, P., Blatt, M., Dedner, A., Engwer, C., Kl\"ofkorn, R., Kornhuber, R.,
  Ohlberger, M., Sander, O.: A generic grid interface for parallel and adaptive
  scientific computing. part {II}: implementation and tests in dune.
\newblock Computing \textbf{82}, 121--138 (2008)

\bibitem{batory2005featuremodels}
Batory, D.: Feature models, grammars, and propositional formulas.
\newblock In: Proc. SPLC, pp. 7--20. Springer (2005)

\bibitem{2012arXiv1209.5145B}
{Bezanson}, J., {Karpinski}, S., {Shah}, V.B., {Edelman}, A.: {Julia: A Fast
  Dynamic Language for Technical Computing}.
\newblock ArXiv e-prints  (2012)

\bibitem{Brandt77}
Brandt, A.: {Multi-Level Adaptive Solutions to Boundary-Value Problems}.
\newblock Mathematics of Computation \textbf{31}(138), 333--390 (1977)

\bibitem{Briggshenson00}
Briggs, W., Henson, V., McCormick, S.: A Multigrid Tutorial, 2nd edn.
\newblock Society for Industrial and Applied Mathematics (SIAM), Philadelphia,
  PA, USA (2000)

\bibitem{bungartz2010pde}
Bungartz, H., Mehl, M., Neckel, T., Weinzierl, T.: The pde framework peano
  applied to fluid dynamics: an efficient implementation of a parallel
  multiscale fluid dynamics solver on octree-like adaptive cartesian grids.
\newblock Computational Mechanics \textbf{46}(1), 103--114 (2010)

\bibitem{Catanzaro:2011:CCE:1941553.1941562}
Catanzaro, B., Garland, M., Keutzer, K.: Copperhead: Compiling an embedded data
  parallel language.
\newblock In: Proceedings of the 16th ACM Symposium on Principles and Practice
  of Parallel Programming, PPoPP '11, pp. 47--56. ACM, New York, NY, USA
  (2011).
\newblock \doi{10.1145/1941553.1941562}.
\newblock \urlprefix\url{http://doi.acm.org/10.1145/1941553.1941562}

\bibitem{christen2011patus}
Christen, M., Schenk, O., Burkhart, H.: Patus: A code generation and autotuning
  framework for parallel iterative stencil computations on modern
  microarchitectures.
\newblock In: Parallel \& Distributed Processing Symposium (IPDPS), 2011 IEEE
  International, pp. 676--687. IEEE (2011)

\bibitem{devito2011liszt}
DeVito, Z., Joubert, N., Palacios, F., Oakley, S., Medina, M., Barrientos, M.,
  Elsen, E., Ham, F., Aiken, A., Duraisamy, K., et~al.: Liszt: a domain
  specific language for building portable mesh-based {PDE} solvers.
\newblock In: Proceedings of 2011 International Conference for High Performance
  Computing, Networking, Storage and Analysis, p.~9. ACM (2011)

\bibitem{Dietrich:2010:AWSEC10}
Dietrich, I., German, R., K\"ostler, H., R\"ude, U.: Modeling multigrid
  algorithms for variational imaging.
\newblock In: Proceedings of 21st Australian Software Engineering Conference
  (ASWEC2010), pp. 224--234. IEEE Computer Society Washington, DC, USA (2010).
\newblock \doi{10.1109/AWSEC.2010.16}

\bibitem{Falgout99}
Falgout, R., Henson, V., Jones, J., Yang, U.: Boomer {AMG}: {A} parallel
  implementation of algebraic multigrid.
\newblock Tech. Rep. UCRL-MI-133583, Lawrence Livermore National Laboratory
  (1999)

\bibitem{Frigo:1999:FFT:301618.301661}
Frigo, M.: A fast fourier transform compiler.
\newblock In: Proceedings of the ACM SIGPLAN 1999 Conference on Programming
  Language Design and Implementation, PLDI '99, pp. 169--180. ACM, New York,
  NY, USA (1999).
\newblock \doi{10.1145/301618.301661}.
\newblock \urlprefix\url{http://doi.acm.org/10.1145/301618.301661}

\bibitem{garshol2003bnf}
Garshol, L.: {BNF and EBNF: What are they and how do they work}.
\newblock acedida pela {\'u}ltima vez em \textbf{16} (2003)

\bibitem{ghosh2010dsls}
Ghosh, D.: {DSLs in action}.
\newblock Manning Publications Co. (2010)

\bibitem{gilboa2004}
Gilboa, G., Sochen, N., Zeevi, Y.: {Image enhancement and denoising by complex
  diffusion processes}.
\newblock IEEE Transactions on Pattern Analysis and Machine Intelligence
  \textbf{26}(8), 1020--1036 (2004)

\bibitem{koestlerapelhistencils13}
Grebhahn, A., Siegmund, N., Apel, S., Kuckuk, S., Schmitt, C., K\"ostler, H.:
  Optimizing performance of stencil code with spl conqueror.
\newblock In: Proceedings of the 1st International Workshop on High-Performance
  Stencil Computations (HiStencils), pp. 7--14 (2014)

\bibitem{Griebel99}
Griebel, M., Zumbusch, G.: Parallel multigrid in an adaptive pde solver based
  on hashing and space-filling curves.
\newblock Parallel Computing \textbf{25}(7), 827 -- 843 (1999)

\bibitem{haase2010parallel}
Haase, G., Liebmann, M., Douglas, C., Plank, G.: A parallel algebraic multigrid
  solver on graphics processing units.
\newblock In: High performance computing and applications, pp. 38--47. Springer
  (2010)

\bibitem{Hackbusch85}
Hackbusch, W.: Multi-Grid Methods and Applications.
\newblock Springer-Verlag, Berlin, Heidelberg, New York (1985)

\bibitem{honigman2006}
Honigman, O., Zeevi, Y.: {Enhancement of Textured Images Using Complex
  Diffusion Incorporating Schroedinger's Potential}.
\newblock In: 2006 IEEE International Conference on Acoustics, Speech and
  Signal Processing, 2006. ICASSP 2006 Proceedings, vol.~2 (2006)

\bibitem{KoestlerRitterHarbin10}
K\"ostler, H., Ritter, D., Feichtinger, C.: {A Geometric Multigrid Solver on
  GPU Clusters}.
\newblock In: D.A. Yuen, L.~Wang, X.~Chi, L.~Johnsson, W.~Ge, Y.~Shi (eds.) GPU
  Solutions to Multi-scale Problems in Science and Engineering, Lecture Notes
  in Earth System Sciences, pp. 407--422. Springer-Verlag, Berlin, Heidelberg,
  New York (2013).
\newblock \doi{10.1007/978-3-642-16405-7_26}.
\newblock \urlprefix\url{http://dx.doi.org/10.1007/978-3-642-16405-7_26}

\bibitem{koestler13HDR}
K{\"o}stler, H., St{\"u}rmer, M., Pohl, T.: Performance engineering to achieve
  real-time high dynamic range imaging.
\newblock Journal of Real-Time Image Processing pp. 1--13 (2013)

\bibitem{kronawitterhistencils13}
Kronawitter, S., Lengauer, C.: Optimization of two jacobi smoother kernels by
  domain-specific program transformation.
\newblock In: Proceedings of the 1st International Workshop on High-Performance
  Stencil Computations (HiStencils), pp. 75--80 (2014)

\bibitem{koestlerkuckukparco13}
Kuckuk, S., Gmeiner, B., K\"ostler, H., R\"ude, U.: A generic prototype to
  benchmark algorithms and data structures for hierarchical hybrid grids.
\newblock In: Proceedings of the International Conference on Parallel Computing
  (ParCo), pp. 813--822 (2013)

\bibitem{fenics:book}
Logg, A., Mardal, K.A., Wells, G.N. (eds.): Automated Solution of Differential
  Equations by the Finite Element Method, \emph{Lecture Notes in Computational
  Science and Engineering}, vol.~84.
\newblock Springer (2012).
\newblock \doi{10.1007/978-3-642-23099-8}.
\newblock \urlprefix\url{http://dx.doi.org/10.1007/978-3-642-23099-8}

\bibitem{MHTKE12a}
{Membarth}, R., {Hannig}, F., {Teich}, J., {K{\"o}rner}, M., {Eckert}, W.:
  Generating device-specific {GPU} code for local operators in medical imaging.
\newblock In: Proceedings of the 26th IEEE International Parallel and
  Distributed Processing Symposium (IPDPS), pp. 569--581. IEEE (2012).
\newblock \doi{10.1109/IPDPS.2012.59}

\bibitem{MHTK12}
{Membarth}, R., {Hannig}, F., {Teich}, J., {K{\"o}stler}, H.: Towards
  domain-specific computing for stencil codes in {HPC}.
\newblock In: Proceedings of the 2nd International Workshop on Domain-Specific
  Languages and High-Level Frameworks for High Performance Computing (WOLFHPC),
  pp. 1133--1138 (2012).
\newblock \doi{10.1109/SC.Companion.2012.136}

\bibitem{odersky2008programming}
Odersky, M., Spoon, L., Venners, B.: Programming in Scala: a comprehensive
  step-by-step guide.
\newblock Artima Inc (2008)

\bibitem{puschel2005spiral}
P{\"u}schel, M., Moura, J.M., Johnson, J.R., Padua, D., Veloso, M.M., Singer,
  B.W., Xiong, J., Franchetti, F., Gacic, A., Voronenko, Y., et~al.: Spiral:
  Code generation for dsp transforms.
\newblock Proceedings of the IEEE \textbf{93}(2), 232--275 (2005)

\bibitem{puschel2004spiral}
P{\"u}schel, M., Moura, J.M., Singer, B., Xiong, J., Johnson, J., Padua, D.,
  Veloso, M., Johnson, R.W.: Spiral: A generator for platform-adapted libraries
  of signal processing alogorithms.
\newblock International Journal of High Performance Computing Applications
  \textbf{18}(1), 21--45 (2004)

\bibitem{Ragan-Kelley:2013:HLC:2491956.2462176}
Ragan-Kelley, J., Barnes, C., Adams, A., Paris, S., Durand, F., Amarasinghe,
  S.: Halide: A language and compiler for optimizing parallelism, locality, and
  recomputation in image processing pipelines.
\newblock In: Proceedings of the 34th ACM SIGPLAN Conference on Programming
  Language Design and Implementation, PLDI '13, pp. 519--530. ACM, New York,
  NY, USA (2013).
\newblock \doi{10.1145/2491956.2462176}.
\newblock \urlprefix\url{http://doi.acm.org/10.1145/2491956.2462176}

\bibitem{sampath2010parallel}
Sampath, R., Biros, G.: A parallel geometric multigrid method for finite
  elements on octree meshes.
\newblock SIAM Journal on Scientific Computing \textbf{32}, 1361 (2010)

\bibitem{SKKHT14}
{Schmitt}, C., {Kuckuk}, S., {K{\"o}stler}, H., {Hannig}, F., {Teich}, J.: An
  evaluation of domain-specific language technologies for code generation.
\newblock In: Proceedings of the 14th International Conference on Computational
  Science and its Applications (ICCSA) (2014)

\bibitem{siegmund2012predicting}
Siegmund, N., Kolesnikov, S., K{\"a}stner, C., Apel, S., Batory, D.,
  Rosenm{\"u}ller, M., Saake, G.: Predicting performance via automated
  feature-interaction detection.
\newblock In: Proceedings of the 2012 International Conference on Software
  Engineering, pp. 167--177. IEEE Press (2012)

\bibitem{tang2011pochoir}
Tang, Y., Chowdhury, R.A., Kuszmaul, B.C., Luk, C.K., Leiserson, C.E.: The
  {P}ochoir stencil compiler.
\newblock In: Proceedings of the 23rd ACM symposium on Parallelism in
  algorithms and architectures, pp. 117--128. ACM (2011)

\bibitem{Trottenbergoosterlee01}
Trottenberg, U., Oosterlee, C., Sch{\"u}ller, A.: Multigrid.
\newblock Academic Press, San Diego, CA, USA (2001)

\bibitem{Whaley00automatedempirical}
Whaley, R.C., Petitet, A., Dongarra, J.J.: Automated empirical optimization of
  software and the atlas project.
\newblock PARALLEL COMPUTING \textbf{27}, 2001 (2000)

\bibitem{Wienands05}
Wienands, R., Joppich, W.: Practical {Fourier} Analysis for Multigrid Methods,
  \emph{Numerical Insights}, vol.~5.
\newblock Chapmann and Hall/CRC Press, Boca Raton, Florida, USA (2005)

\end{thebibliography}
%

\end{document}